\documentclass{cimento}

\setcopyright{The Author(s)}
\usepackage{graphicx}  
\usepackage{amsmath}
\usepackage{verbatim}
\usepackage{xcolor}
\usepackage[font=small]{caption}
\usepackage{subcaption}

\title{Realization and characterization of a low intensity noise ultrafast Yb-doped fiber amplifier}

\author{F.~Canella\from{ins:p}\from{ins:i}\ETC,
on behalf of
L.M.~Molteni\from{ins:p}\from{ins:c},
S.~Cialdi\from{ins:s}\from{ins:i},
P.~Laporta\from{ins:p}\from{ins:c},
N.~Coluccelli\from{ins:p}\from{ins:c}
        \atque
G.~Galzerano\from{ins:c}\from{ins:p}
}
\instlist{\inst{ins:p} Dipartimento di Fisica, Politecnico di Milano - Milano, Italy
\inst{ins:i} INFN, Sezione di Milano - Milano, Italy
\inst{ins:c} Istituto di Fotonica e Nanotecnologie-CNR - Milano, Italy
  \inst{ins:s} Dipartimento di Fisica, Universit\`a degli Studi di Milano - Milano, Italy}

\begin{document}

\maketitle

\begin{abstract}
We report on the design and whole characterization of low-noise and affordable-cost Yb-doped double-clad fiber amplifiers operating at room temperature in the near-infrared spectral region at pulse repetition rate of 160 MHz. Two different experimental configurations are discussed.
In the first one, a broadband seed radiation with a transform limited pulse duration of 71~fs, an optical spectrum of 20~nm wide at around 1040~nm, and $\sim$20~mW average power is adopted. In the second configuration, the seed radiation is constituted by stretched pulses with a time duration as long as  170 ps, with a 5-nm narrow pulse spectrum centered at 1029 nm and $\sim$2 mW average input power. In both cases we obtained transform limited pulse trains with an amplified output power exceeding 2 W. Furthermore, relative intensity noise measurements show that no significant noise degradation occurs during the amplification process.
\end{abstract}

\section{Introduction}\label{intro}
In the last decades, high-peak power and low-noise optical pulses with high-repetition rates (larger than 100~MHz) in the near-infrared spectral region gained attention for their multiple applications in several fields of industrial, defense, and scientific applications \cite{ref:gen1,ref:gen2,ref:gen3}. Low intensity noise pulses can be generally obtained using low-power oscillators with high-repetition pulse rate and high quality factors.
To further increase the average output power without intensity degradation, low intensity noise ultrafast optical amplifiers
have to be developed \cite{ref:sido}.\\
In this framework, ytterbium-doped fibers are good candidates as amplifiers gain media, thanks to their ability to amplify radiation over the very broad wavelength range (from 975 to 1200 nm) \cite{ref:paschotta}. 
Furthermore, Yb$^{3+}$ shows a quite simple energy structure, resulting in a quasi-three-level scheme. Indeed, in the 4f$^{13}$ electronic configuration only two level manifolds are relevant: $^2$F$_{7/2}$ and $^2$F$_{5/2}$. The first one acts as ground state, while the second as excited state. Both manifolds are split respectively into four and three Stark sub-levels \cite{ref:valles}. The energy gap between them belongs to the near infrared spectral region, while upper energy levels are in the deep-ultraviolet range. Therefore, an Yb amplifier does not present excited-state absorption and many others deleterious energy transfer processes found in systems based on different dopants such as neodymium (Nd$^{3+}$), erbium (Er$^{3+}$) and thulium (Tm$^{3+}$) \cite{ref:wu}.
To increase the amplifier efficiency, innovative fiber designs and pumping configurations have been studied in the last decades. In particular, in the cladding pump scheme the pump is launched in the fiber cladding instead of the doped core \cite{ref:sn}.
In this way, the pump is gradually absorbed by the active core, but propagates in widely larger region, allowing the use of high-power multimode diode lasers. This approach enhances the pump coupling efficiency, while alignment becomes uncritical.
In the next sections we will present compact, low cost cladding-pumped ultrafast Yb-doped fiber amplifiers operating at room temperature.
Firstly, sect.~\ref{equations} shows the most relevant equations for Yb-fiber amplifiers whereas the next sects.~\ref{setup} and \ref{performances} report on the designed experimental setups and the complete characterization of our amplifiers, respectively. Then we summarize the results in the conclusions section.

\section{Amplifier equations}\label{equations}
The amplifier we present in this paper is based on a double-clad Yb-doped fiber.
As mentioned above, these fibers behave as quasi-three-level systems whose emission and absorption cross sections and energy levels are shown in fig. 1.
\begin{center}
\label{1}
\includegraphics[width=0.95\linewidth]{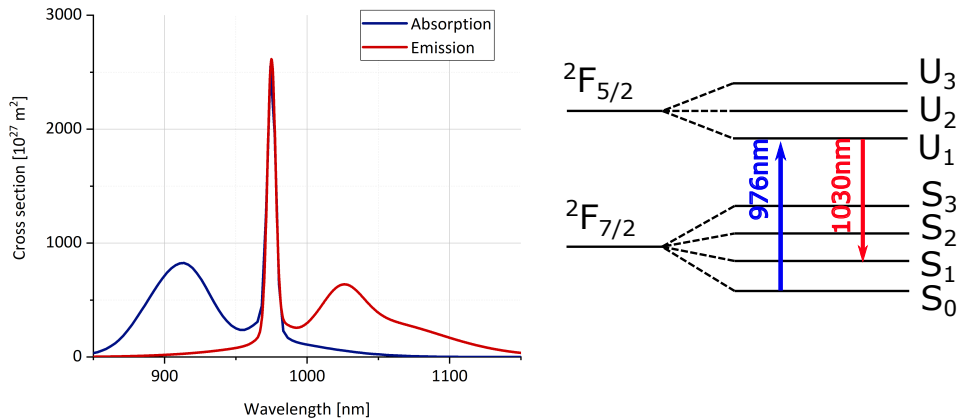}
\captionof{figure}{Left, Yb-doped silica fiber emission and absorption cross sections \cite{ref:paschotta}. Right, quasi-three level structure of Yb$^{3+}$: the pump radiation at 970~nm excites the transition from the Stark sub-level $S_{0}$ to $U_{1}$. The amplification occurs when from such sub-level the electrons radiatively decay on the $S_{1}$ sub-level generating a photon at a wavelength of 1030~nm.}
\end{center}
Here the ground state is the sub-level $S_0$, the excited state is the sub-level $U_1$ and from there ions decay on level $S_1$. The transition $S_1$-$S_2$ occurs by a fast non-radiative process, so it is possible to write the total density of Yb$^{3+}$ ions as $N_{TOT}=N_1+N_2$, where $N_1$ and $N_2$ are respectively the $S_0$ and $U_1$ populations, neglecting $S_1$.
Given this assumption, the amplification process in the fiber is governed by the following spatio-temporal equations \cite{ref:paschotta}.\\
\begin{equation}\begin{split}
\frac{dN_2(t,z)}{dt}=-\frac{dN_1(t,z)}{dt}&=
-\left[\left(\frac{\lambda_p\Gamma_p\sigma_{e}\left(\lambda_p\right)}{hcA_{core}}
\right)P_p+\left(\frac{\lambda_s\Gamma_s\sigma_{e}\left(\lambda_s\right)}{hcA_{core}}\right)P_s\right]N_2+\\
&+
\left[\left(\frac{\lambda_p\Gamma_p\sigma_{a}\left(\lambda_p\right)}{hcA_{core}}\right)P_p+\left(\frac{\lambda_s\Gamma_s\sigma_{a}\left(\lambda_s\right)}{hcA_{core}}\right)P_s\right] N_1
-\frac{N_2}{\tau};
\end{split}
\end{equation}
\begin{equation}
\frac{dP_P(z)}{dz}=\Gamma_p\left(\sigma_{e}\left(\lambda_p\right)N_2-\sigma_{a}\left(\lambda_p\right)N_1\right)P_p(z);
\end{equation}
\begin{equation}
\frac{dP_S(z)}{dz}=\Gamma_s\left(\sigma_{e}\left(\lambda_s\right)N_2-\sigma_{a}\left(\lambda_s\right)N_1\right)P_s(z).
\end{equation}
In eq.1, 2, 3 the subscripts \textit{p} and \textit{s} indicate respectively the pump and the seed signal, $\sigma_e$ and $\sigma_a$ are the emission and absorption cross sections (fig. 1), $\tau$ represents the spontaneous decay time from the excited energy level (tipically $\approx0.8$ ms), $\lambda_p$ and $\lambda_s$ are the pump and seed wavelengths, while $\Gamma_p$ and $\Gamma_s$ are the spatial overlap factors. These latter elements are particularly important for double-clad fibers, where the Yb$^{3+}$ ions are present only inside the core and not in the cladding. For this reason, only a part of the pump profile (guided in the cladding) interact with the doping. At the same time, the seed radiation is not perfectly confined in the core area $A_{core}$, so a small correction is needed. Assuming a step-index fiber profile, a flat-top pump beam and a gaussian mode for the seed, the overlap factors are $\Gamma_s=1-\exp(-2A_{core}/A_{mode})$ and $\Gamma_p=A_{core}/A_{cladding}$ \cite{ref:barnard}.
Equations 1, 2 and 3 can be numerically solved in the steady-state condition and allow to simulate the amplifier performances for different fiber parameters and coupled power.\\
The equations above describe a quite simple model, but in certain cases a more complete description may be required. In particular, the Amplified Spontaneous Emission (ASE) should be taken in account in specific situations, as well as ions quenching or radiative emission by Yb$^{3+}$ dimers, trimers and tetrames (especially in strongly-doped fibers). Examples of such complex equations can be found at references \cite{ref:wu} and \cite{ref:barnard}.

\section{Experimental setup}\label{setup}
The optical amplifiers we developed are based on 1.8 m-long polarization maintaining highly-doped double-clad active fiber (Liekki Yb1200-10/125DC-PM) with a core diameter of 10~$\mu$m, a pump cladding diameter of 125~$\mu$m, and a peak cladding absorption of 6.5~dB/m at 976~nm pump wavelength.
The active fiber is pumped by a commercially available laser CW diode at 976~nm with 6~W of power in forward configuration, while the seed pulses are generated by the Yb:CaGdAlO \cite{ref:pirzio} SESAM mode-locked source at a central wavelength of 1040~nm with a duration of 71~fs, 160-MHz repetition rate, and an average power of 50~mW.\\
\begin{center}
\label{f2}
\includegraphics[width=0.7\linewidth]{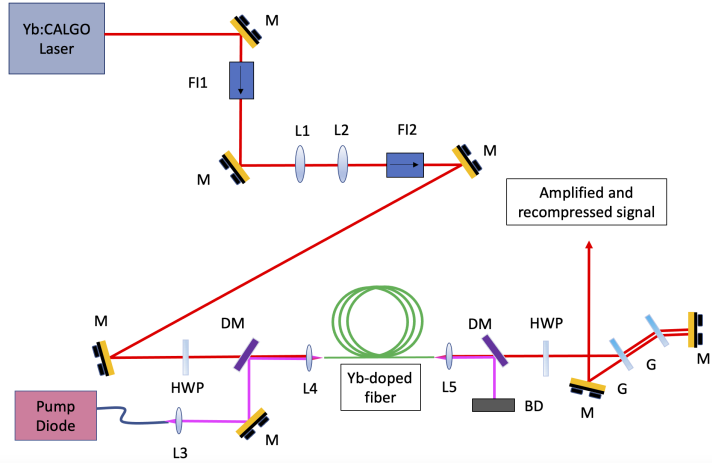}
\captionof{figure}{Experimental setup of the unstretched pulses configuration. M: mirror; FI: Faraday isolator; L: lens; HWP: half-wave plate; DM: dichroic mirror; BD: beam dumper; G: grating.}
\end{center}

The first experimental configuration we realized and studied is sketched in fig. 2.
At the laser output a low power Faraday isolator (FI1) is positioned to protect the laser from dangerous reflections.
Since the launch in the Yb-fiber of the seed and the pump are performed by the same lens (L4), their beam diameters must be adjusted to reach the correct dimension inside the fiber. To this aim, a telescope made of two lenses, L1 and L2, is positioned after FI1.
The focal lengths of  the lenses are 25~mm and 70~mm respectively and the magnification factor of the telescope is about 2.8 for a distance between the lenses equal to 95~mm. In this way, the beam diameter -initially 1.4 mm- becomes 2 mm, namely the value we desire for an optimal launch in the Yb fiber using L4.
After the telescope, there is another high-power Faraday isolator (FI2). This is needed to avoid positive feedbacks in the amplifier caused by the telescope lenses which could be dangerous for the amplifier itself because they could lead to unwanted parasitic lasing.
Then the beam is sent to a couple of golden mirrors, which are useful to manage the beam tilt. Before the launch there are also an half-wave plate and a dichroic mirror (HWP and DM). The HWP manages the seed polarization and to tune it according to the polarization maintaining axis of the doped fiber, while the dichroic allows the alignment between the seed and the pump -after the collimation via L3-.
After the alignment on the dichroic mirror, both pump and seed are launched in fiber by a 30 mm-focal length lens (L4).
The alignment between the lens and the fiber is performed by a  3D movable mount where the lens L4 is fixed on. A similar component is used for the output collimation for L5.\\
At the fiber input, the seed power is 14 mW, since part of it has been lost by the Faraday isolators, the mirrors and the lenses. On the other hand, the pump diode power can easily reach tens of Watts, but  it is limited only up to 6 W to avoid fiber damaging.
At the fiber output there is a collimation lens (focal length of 19 mm) and a dichroic mirror useful to separate the amplified seed from the residual pump. Then an half-wave plate is positioned to control the polarization before the recompression stage.
The recompression stage is needed to compensate for the normal group velocity dispersion of the active fiber. At the fiber output the measured pulse length is 1.66 ps against the input pulse duration of 71~fs.\\
A couple of transmission gratings (Wasatch Photonics, 800 grooves/mm) at a distance of 1.12 cm at 24 degrees allows a complete recompression to 71 fs in a double passage, with a power efficiency of 55$\%$. 
\begin{center}
\label{f2}
\includegraphics[width=0.7\linewidth]{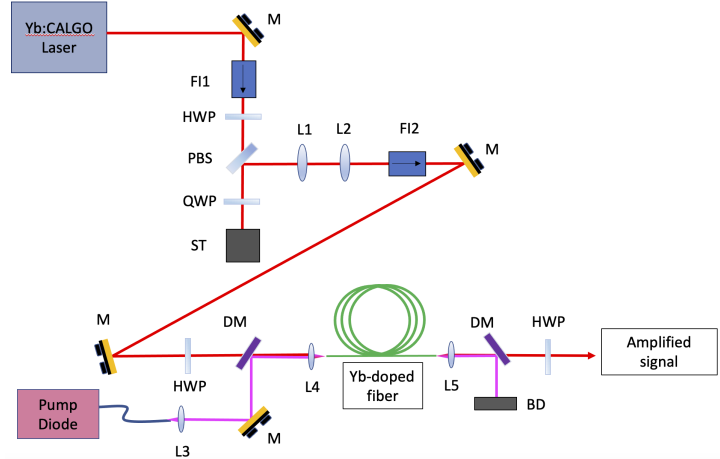}
\captionof{figure}{Experimental setup of the stretched pulses configuration. M: mirror; FI: Faraday isolator; HWP: half-wave plate; QWP: quarter-wave plate; ST: VBG stretcher; L: lens; DM: dichroic mirror; BD: beam dumper.}
\end{center}

We also studied an experimental configuration in which the laser pulses are filtered in bandwidth and stretched in time before fiber launch (fig. 3). Such kind of systems are preferable when avoiding nonlinear effects during the amplification is strictly mandatory.
To stretch the pulses, we use a Volume-Bragg-Grating (VGB) stretcher: it is an optical component in which different wavelengths are reflected at different depths inside it, providing a chirp in frequency and a temporal stretching that allow to reach a time duration of 170 ps.
It is worth notice that due to the VGB small spectral bandwidth - 5 nm centered at  1029 nm - only a part ($\sim$10$\%$) of the laser spectrum is selected and stretched. In this way, also a part of the seed power is lost before the Yb fiber, where only 2.4 mW are coupled.
As shown in fig. 3, the VBG is positioned after the first Faraday isolator. A system made by a half-wave plate, a quarter-wave plate and a Polarizing Beam Splitter is employed to deviate the pulses after the VGB reflection and temporal elongation.
Then, the laser pulses follow the same optical path of the previous setup, except for the recompression stage. In principle, it is possible to recompress the pulse after the amplification process with a second VGB, but we simply demonstrated the amplification without recompression. 

\section{Amplifier performances}\label{performances}

The more relevant experimental results are exposed in the following section.
In particular, for both the configurations we report  the amplifier output power as a function of the pump power, the gain and the output power dependency on the seed power (at fixed pump power), the pulses spectra and - only in the unstretched pulses case - a comparison between the autocorrelation of the laser pulses and the amplified pulses after the recompression process. Finally, Relative Intensity Noise (RIN) measurements are presented.
Power characterizations have been performed by a commercially available thermal power sensor (Coherent LM10), the RIN data have been acquired using an electrical spectrum analyzer (Agilent E445A) reading a signal from a fast InGaAs avalanche photodetector (Thorlabs APD430C/M), while the pulse length have been measured with a non-collinear second-harmonic autocorrelator (Femtochrome FR103XL).

Firstly, we characterized the output spectra for all the experimental configurations. Figure 4 shows the recorded output spectra together with the spectra of the input seeder. In particular the full laser spectrum  (FWHM=22 nm) is drawn on the left, while the spectrum of stretched pulses (FWHM=5 nm) is represented on the right .
A partial deformation in spectra profiles is noticeable and a small shift in wavelength for the unstretched pulse can ben observed.
The blueshift is explained by a higher value of the emission cross section (see fig. 1) at 1030 nm than at 1040 nm, which advantages the amplification of spectral components at  lower wavelengths resulting in a shift of the whole spectrum profile. Regarding the deformation effects, they are due to nonlinear effects in fiber, \textit{e.g.} self phase modulation \cite{ref:agrawal}. Nonlinear phenomena are more evident in the unstretched configuration, where the pulses peak power is higher than in the stretched case. Nevertheless, nonlinear effects do not compromise the pulses quality and the conserved symmetry is largely enough to fully recompress the pulses.
\begin{center}
\includegraphics[width=0.49\linewidth]{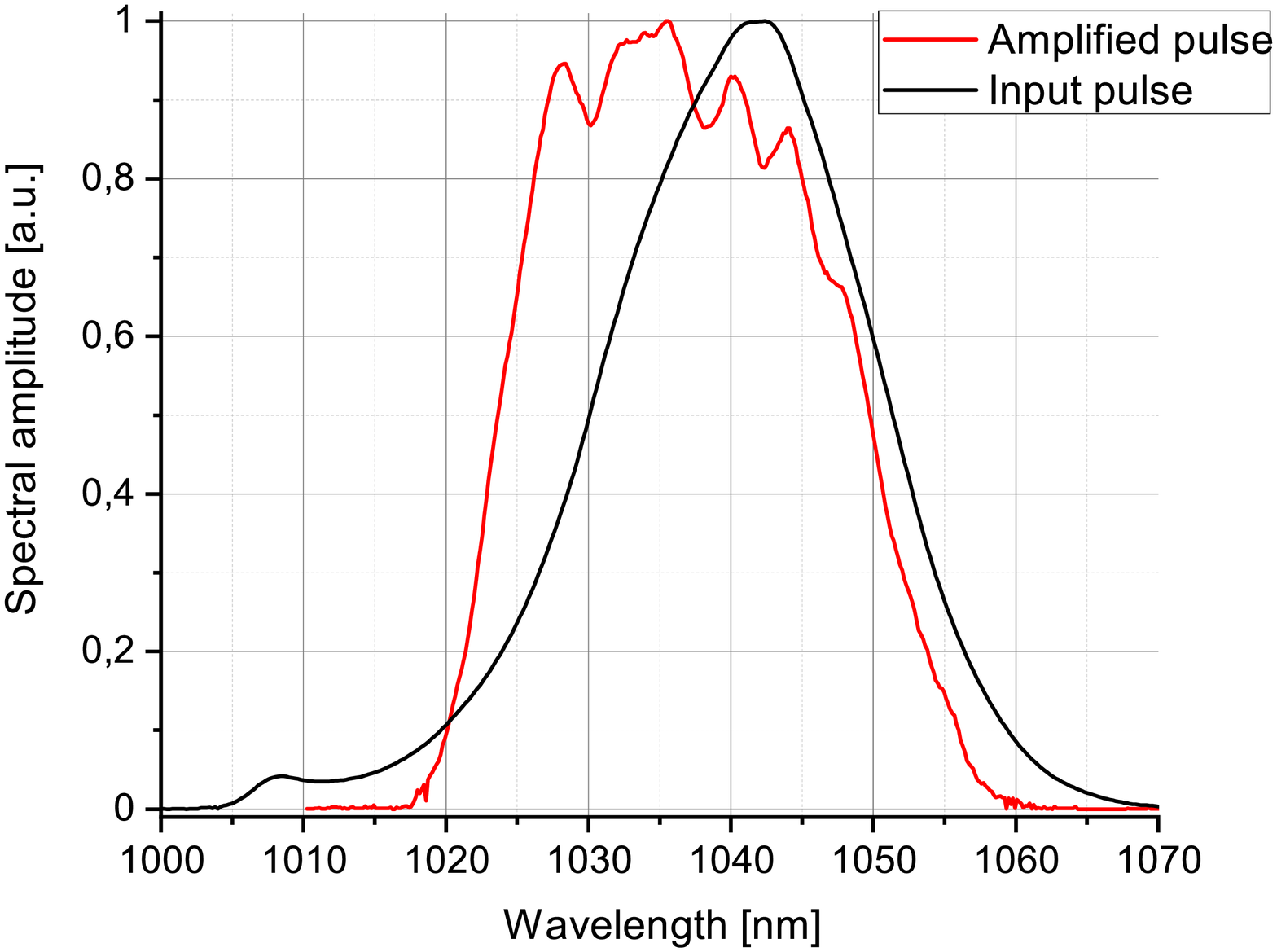}
\includegraphics[width=0.49\linewidth]{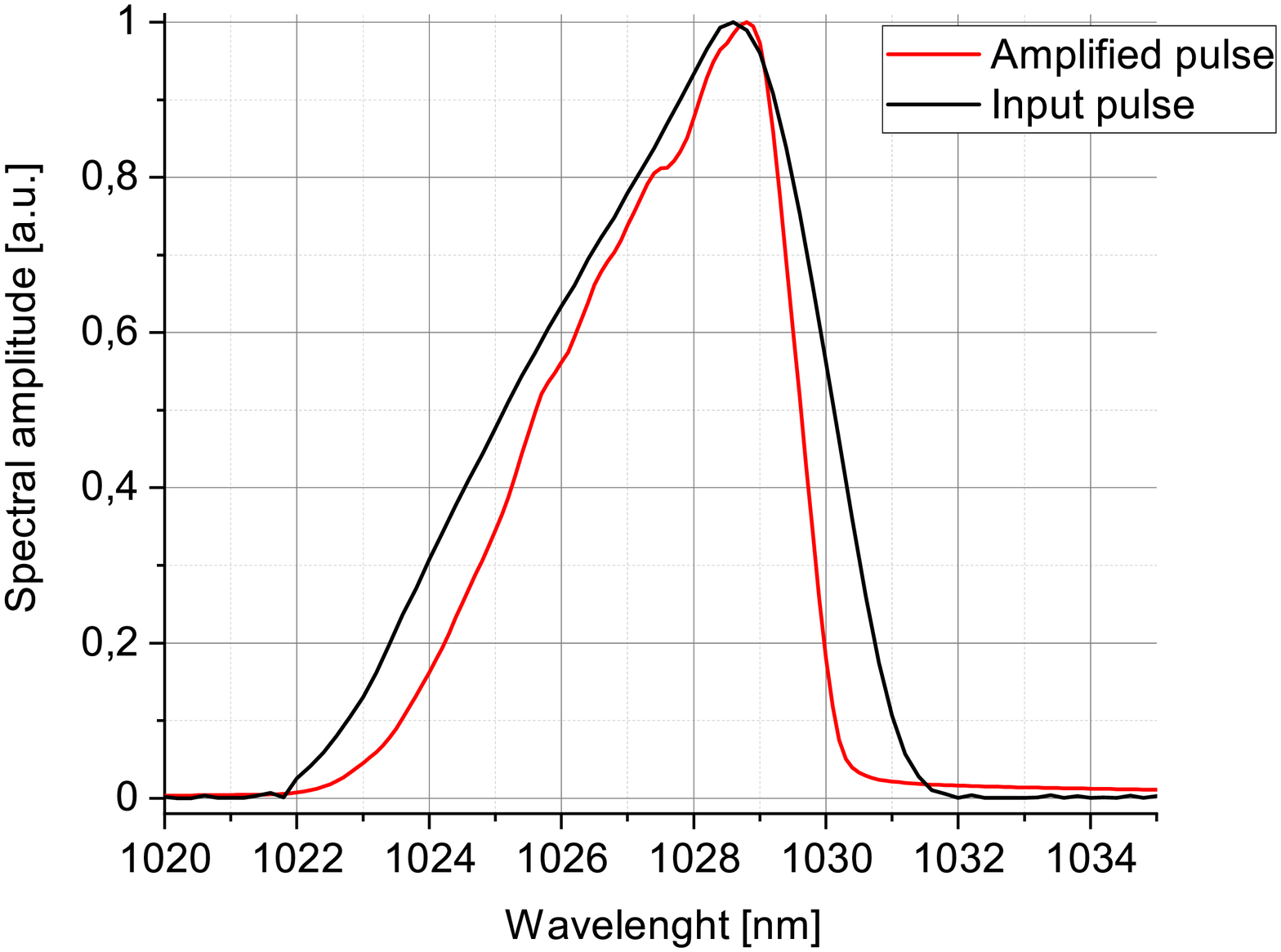}
\captionof{figure}{Left, amplified pulses spectrum in the unstretched configuration. Right, amplified pulses spectrum in the stretched configuration.}
\end{center}
The effectiveness of the recompression process is demonstrated in fig. 5, where both the input seed and amplified pulses autocorrelations are plotted. The gratings pulse compressor compensate the dispersion broadening experienced by the pulses passing through the Yb fiber and the other optical elements in the experimental setup. The measured autocorrelation FWHMs are both equal to 109 fs. Successively, the FWHMs are divided by 1.54, because the pulses shape has been approximated to a sech$^2$ \cite{ref:treacy}. As a result, the two measured pulses are $\Delta\tau_{\textit{pulses}}$=71 fs. Anyway, a small third-order contribution is visible in the recompressed pulses autocorrelation, as symmetrical wings on the sech$^2$ tails. However, such contribution was not relevant to our purposes, because we are still able to retrieve the initial pulse duration.
\begin{center}
\includegraphics[width=0.6\linewidth]{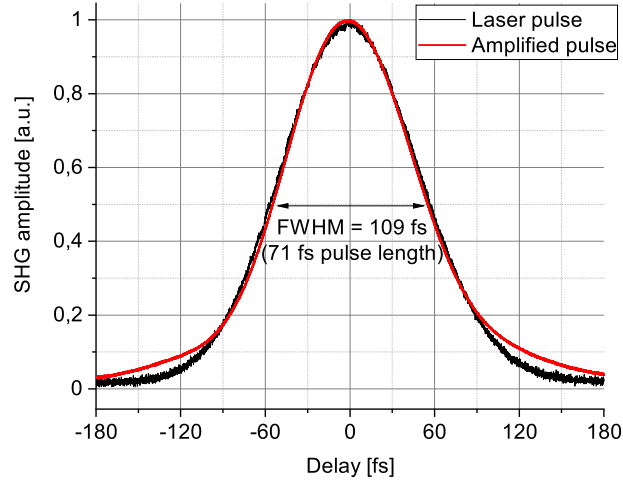}
\captionof{figure}{Intensity autocorrelations of the laser output and after the recompression. In both the measures the  pulse temporal length is 71 fs.}
\end{center}

For what concerns the power performances characterization, figs. 6 and 7 give a complete picture both in terms of output power and optical gain.\\
Figure 6 shows the power and gain curves for the unstretched configuration. Looking at fig. 6a (on the left) it can be noticed that the output power increases approximately linearly with the pump power. The maximum  provided pump power is 6 W (no higher power has been  tested, in order to avoid damages to the doped fiber), corresponding to a maximum output power of 2.6 W. With a seed input power of 14 mW, the amplifier gain is 22.7 dB. 
However, higher amplification gain is reached when the seed power is lower. The complete gain plot is drawn in fig. 6 on the right, as well as the output power as a function of the seed power.
The input power has been modulated by a variable neutral density filter keeping the pump power fixed to 5.4 W, then the output power and the amplifier gain have been acquired.
Even though the output power decreases when the seed power is lower, the amplifier gain enhances to 32.3 dB when the seed power is 1 mW, showing a saturating behavior when the seed power increases.
\begin{center}
\label{prova4}
\includegraphics[width=0.47\linewidth]{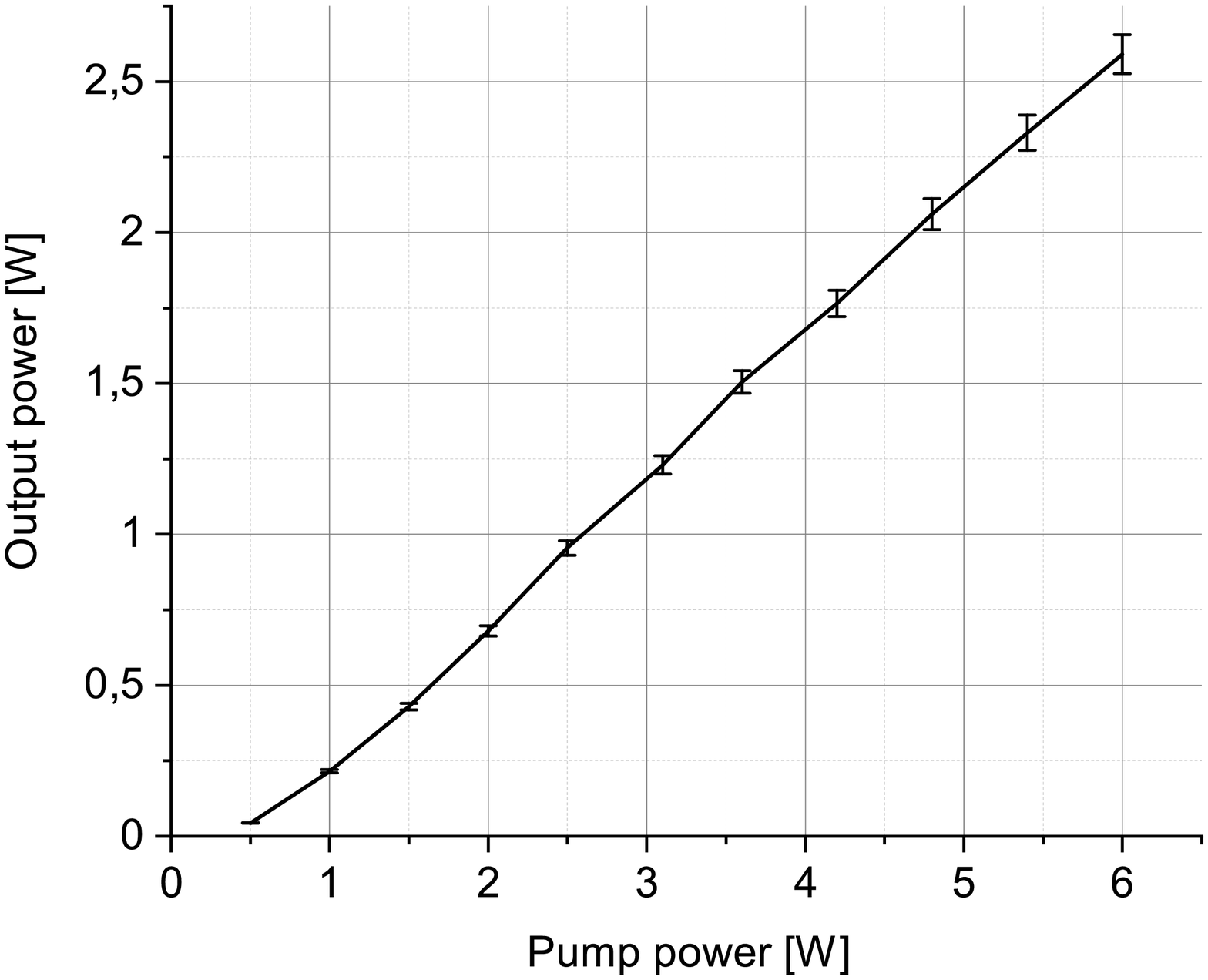}
\includegraphics[width=0.51\linewidth]{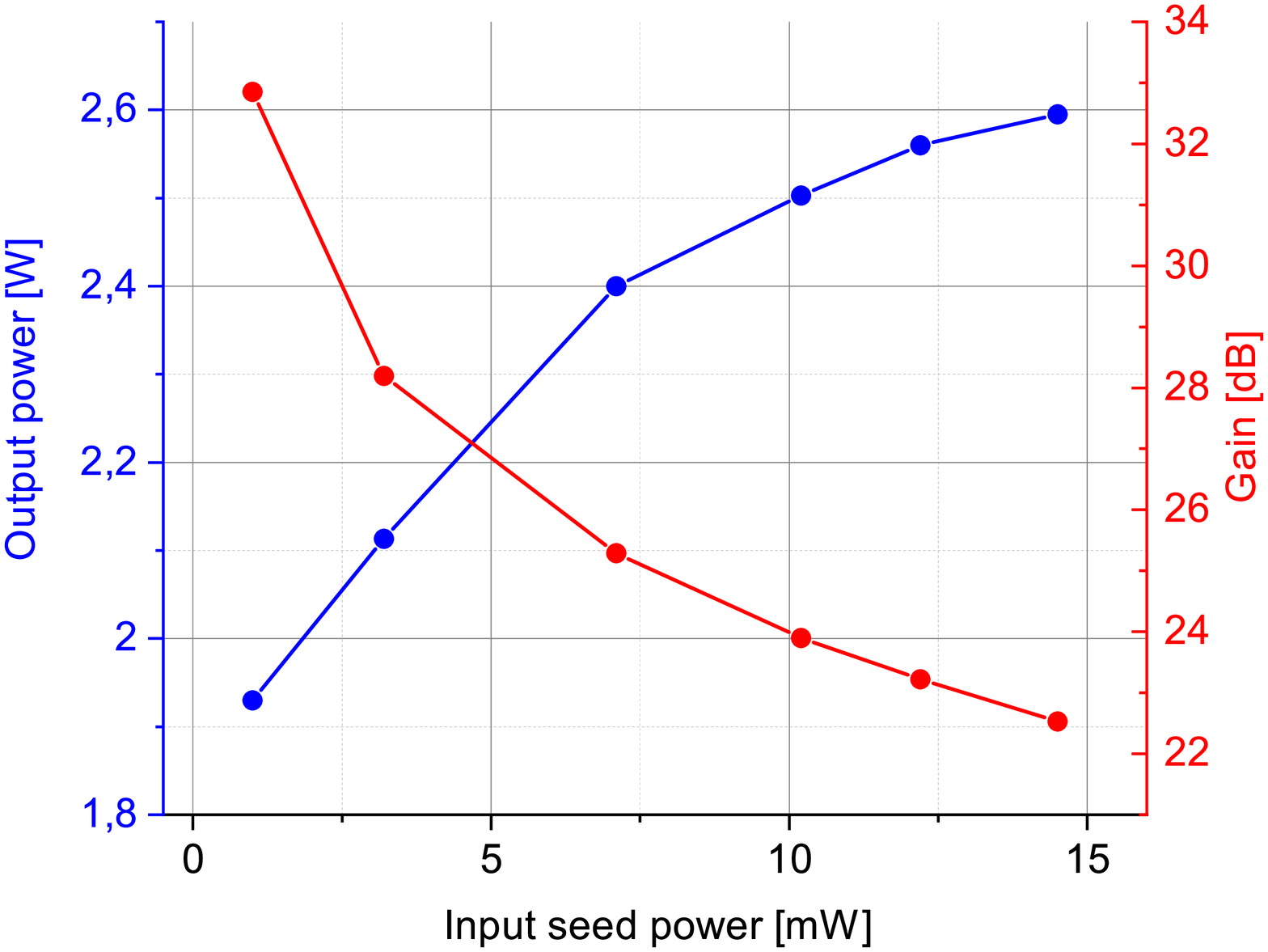}
\captionof{figure}{Unstretched pulses configuration. Left, amplifier output power as a function of the pump power. Seed average power maintained to 14 mW. Right, amplifier output power (left axis) and gain (right axis) as a function of the seed power. Pump power maintained to 5.4 W.}
\end{center}
\begin{center}
\includegraphics[width=0.48\linewidth]{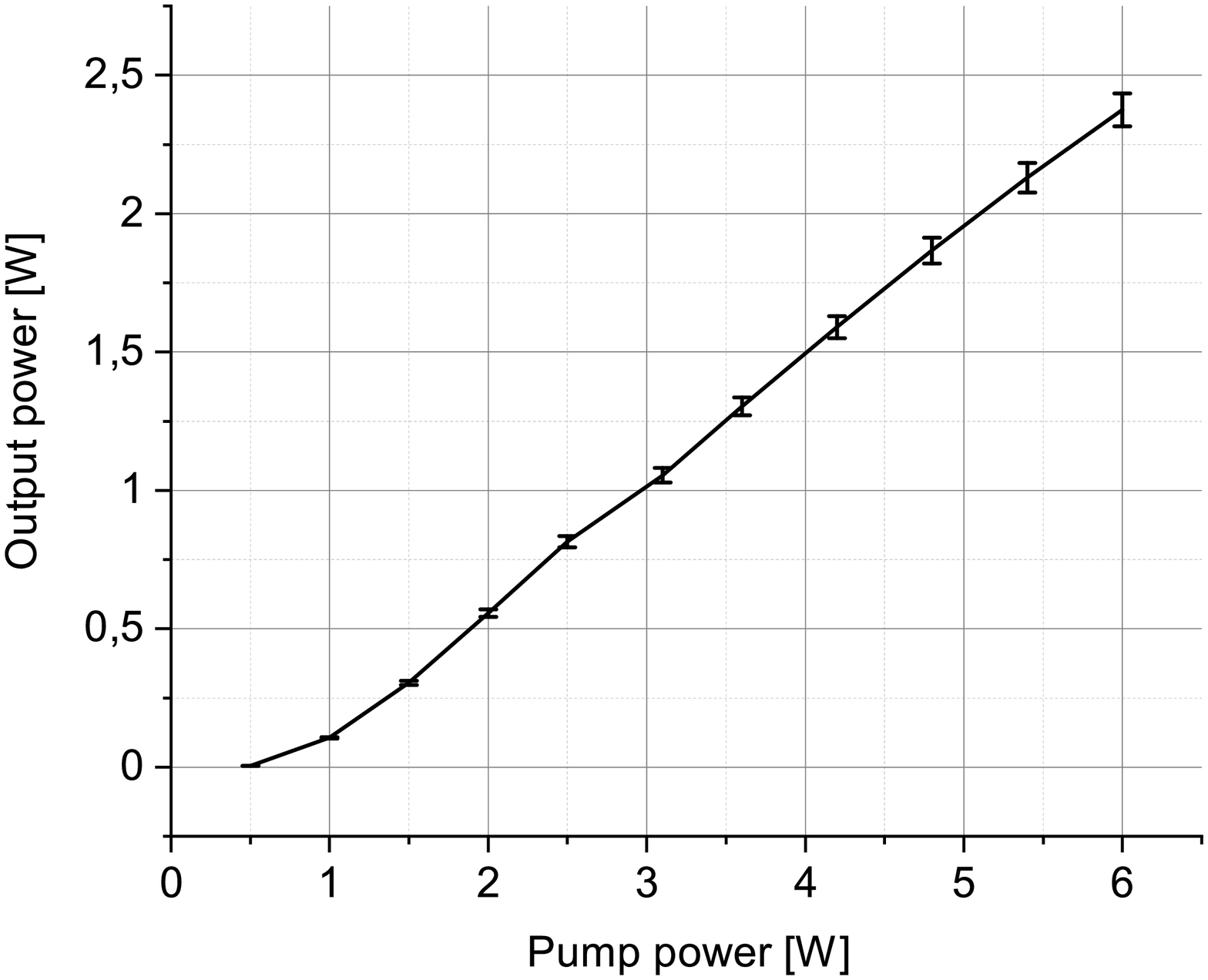}
\includegraphics[width=0.51\linewidth]{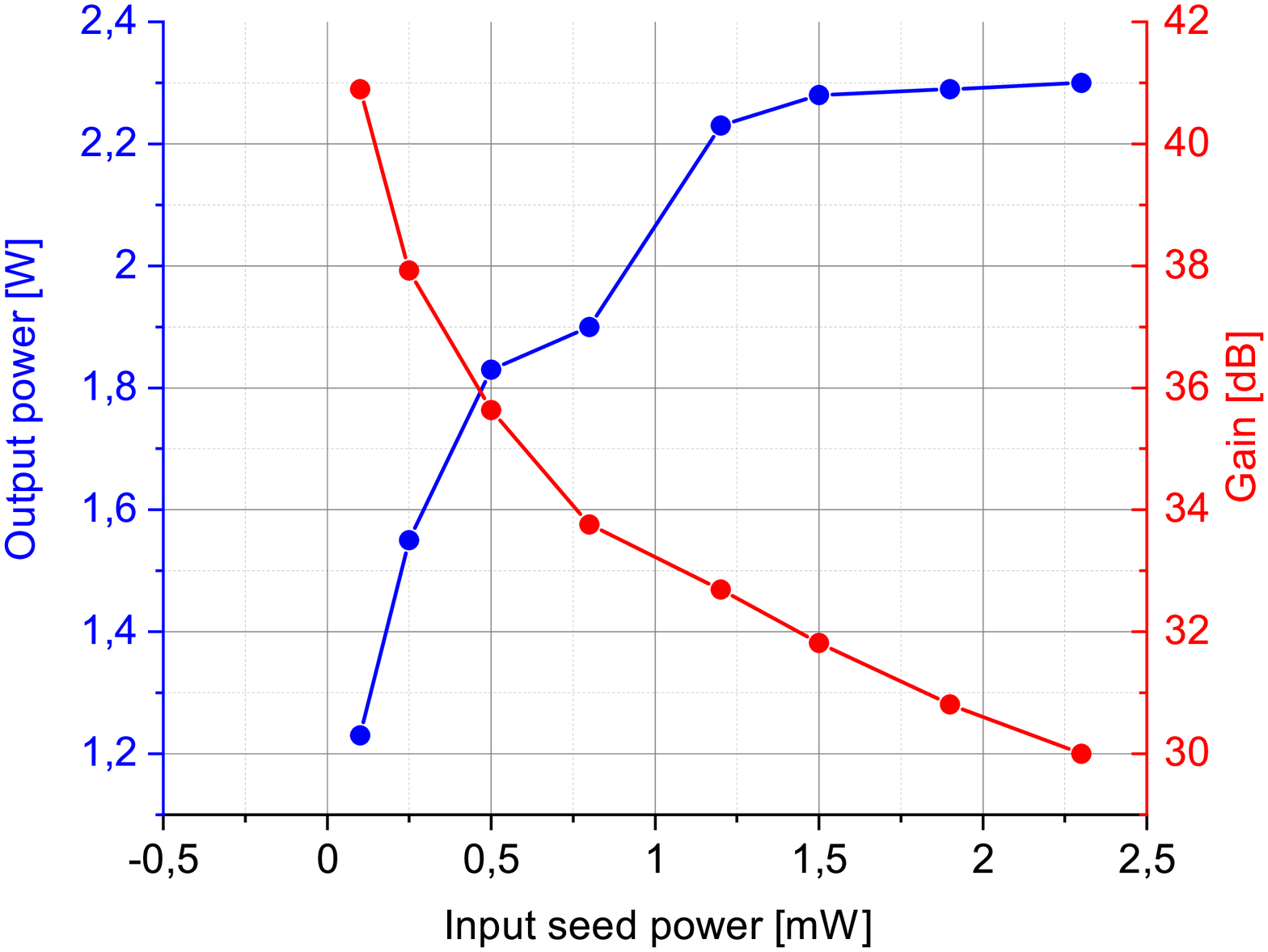}
\captionof{figure}{Stretched pulses configuration. Left, amplifier output power as a function of the pump power. Seed average power maintained to 2.4 mW. Right, amplifier output power (left axis) and gain (right axis) as a function of the seed power. Pump power maintained to 5.4 W.}
\end{center}

The same measurements are performed for the stretched pulses configuration as reported in fig. 7.
Although we find substantially the same behavior for output power, some differences in the amplifier gain can be found.
In fact, fig. 7 shows on the left that the maximum output power is 2.4 W, which is quite similar to the highest power in fig. 6a, although the seed power is six times lower (2.4 mW). Looking at fig. 7b (right) we find that in this case the gain is 41dB and also for lower seed power values the gain remains  approximately one order of magnitude higher than in the unstretched case.
From these plots it is clear that the amplification process is more efficient for stretched laser pulses due to the narrower pulse bandwidth of the configuration with respect to the unstretched one.\\
In addition, it is worth noticing that there is a saturating trend of the output power in both figs. 6b and 7b. For the stretched pulses curve such behavior is more evident: when the seed power becomes higher than 1.2 mW the curve clearly flattens.
It is worth noting that operating in saturation regime gives important benefits in terms of signal noise degradation during the amplification.
Such advantages have been described by J. Zhao \textit{et al.} in ref.\cite{ref:zhao}. Consequently, it is crucial to maintain the seed power close to the saturation to keep a low intensity noise of the amplified pulses. Figure 8 shows the measured RIN spectra of the amplifier pump diode, of the seed laser, and of the amplified pulses. Even if the amplifier pump noise is on average 20 dB higher than the laser source noise, it doesn't degradate the signal outgoing the saturated amplifier, that remains low in all the studied configurations. All RIN spectra refer to a pump power equal to 6 W.
Only a couple of low-frequency noise peaks (100 Hz and 110 Hz) are transferred from the pump to the amplified seed, but their contribution on the considered bandwidth is quite negligible.
Aside this exception, the seed traces before and after the amplification describe the same decreasing trend from -70 dB/Hz at 1 Hz to less than -100 dB/Hz around 10 kHz and then they maintain this low value before approaching the floor level at 10$^5$ Hz.

\begin{center}
\includegraphics[width=\linewidth]{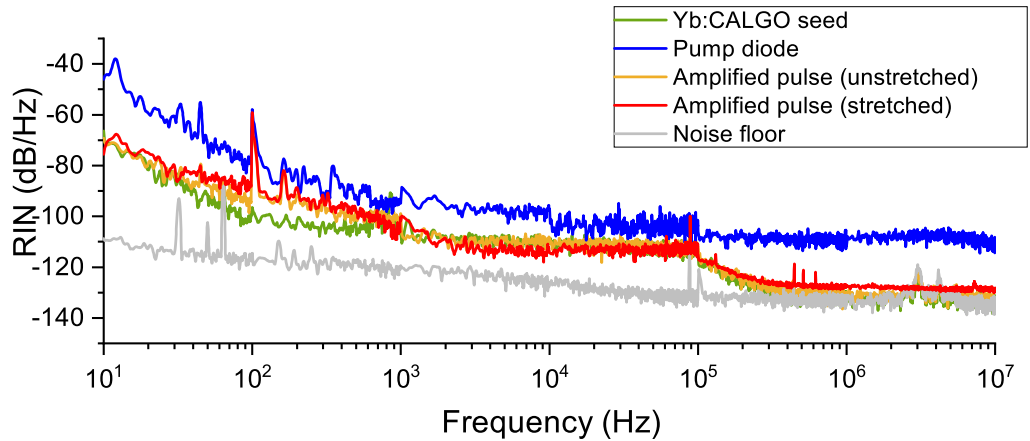}
\captionof{figure}{Relative Intensity Noise measures on the seed signal, on the pump signal and on the amplified signal in both stretched and unstretched configuration. Pump power is fixed to 6 W.}
\end{center}

\section{Conclusions}\label{conclusions}
In this paper we described the design and characterization of a compact and low cost double-clad Yb-doped fiber amplifier, testing its performances in two different experimental configurations. For each setup, amplified power, spectra, optical gain, and RIN measurements have been presented.
In both configurations low intensity noise pulses have been demonstrated with average output power higher than 2 W and transform limited pulse duration as short as 71 fs. In conclusion, we provided a low cost and powerful setup suitable for a wide variety of ultrafast optical systems and experiments.

\acknowledgments
The authors acknowledge INFN Milano, IFN-CNR and Politecnico di Milano for the given financial support.

\end{document}